\documentclass{PoS}
\usepackage{mathrsfs}
\usepackage{subfigure}
\usepackage{verbatim}

\def\mpt{{\slash\!\!\!\!\!\:P}_T}

\def\mptv{{\slash\!\!\!\!\!\:\vec{P}}_T}

\title{$Z'_{B-L}$ discovery potential at the LHC for $\sqrt{s}=7$ TeV}

\ShortTitle{$Z'_{B-L}$ discovery potential at the LHC for $\sqrt{s}=7$ TeV}

\author{\speaker{L.~Basso}$^{,1}$, A.~Belyaev$^1$, S.~Moretti$^{1}$, G.M.~Pruna$^1$ and C.H.~Shepherd-Themistocleous$^{1,}$\thanks{This work is supported in part by the SEPnet Physics Consortium.}\\
        $^1$NExT Institute, University of Southampton, Highfield, Southampton, 
        and RAL, Chilton, Didcot, ~UK\\
        E-mail: \email{lb4x07@soton.ac.uk}, ~\email{a.belyaev@soton.ac.uk}, \email{s.moretti@soton.ac.uk},~\email{g.m.pruna@soton.ac.uk},~\email{claires@mail.cern.ch}}

\abstract{We present the Large Hadron Collider (LHC) discovery
potential in the $Z'$ sector of a $U(1)_{B-L}$
enlarged Standard Model also encompassing three heavy Majorana
neutrinos, for $\sqrt{s}=7$ centre-of-mass energy, considering both the $Z'_{B-L}\rightarrow e^+e^-$ and $Z'_{B-L}\rightarrow \mu^+\mu^-$ decay channels.

Electrons provide a higher sensitivity to smaller couplings at small $Z'_{B-L}$ boson masses than do muons.
The run of the LHC at $\sqrt{s}=7$ TeV, assuming at most $\int 
\mathcal{L} \sim 1$ fb$^{-1}$, will be able to give similar results to those that will be available soon at the Tevatron in the lower mass region, and to extend them for a heavier $M_{Z'}$. A $5\sigma$ discovery could be possible up to $M_{Z'}=1.2(0.9)$ TeV at the LHC(Tevatron), while a $2\sigma$ exclusion at the LHC could be possible up to  $M_{Z'}=1.6$ TeV.
The new gauge coupling $g'_1$ can been probed, at $5\sigma$, down to $\sim (3 \div 4) \cdot 10^{-2}$ with electrons and down to $\sim (4 \div 5) \cdot 10^{-2}$ with muons, both at the LHC and at the Tevatron, for $M_{Z'}=600$ GeV.

The $Z'$ boson in this model exhibits novel signatures at the LHC, 
as multi-lepton and multi-jet decays via heavy neutrinos, that allow one to measure the heavy neutrino masses involved. Lastly, the simultaneous
measurement of both the heavy neutrino mass and decay length (over a large region of parameter space, the heavy neutrinos are rather long-lived particles) enables an estimate of the absolute mass of the parent light neutrino.}

\FullConference{35th International Conference of High Energy Physics\\
                 July 22-28, 2010\\
                 Paris, France}

\begin{document}
\baselineskip=13.8pt
\section{Introduction}
The 
$B-L$ (baryon number minus lepton number) symmetry plays an important role in
various physics scenarios in and beyond the Standard Model (SM). First, the accidental $U(1)_{B-L}$ global symmetry is not anomalous in the SM with massless neutrinos, but its origin is not well understood.
Second,  its gauged version is 
contained in {several} Grand Unified Theories.
%, {such as the one based on the} $SO(10)$
% group~\cite{Buchmuller:1991ce} (which can have an intermediate symmetry
% ${SU(2)}_L \times {SU(2)}_R \times {SU(3)}_C \times {U(1)}_{B-L}$ 
%which breaks down to the group of the SM at a scale of $10^{10-12}$ GeV).
Third, the scale of  
the $B-L$ symmetry  breaking is related to the mass scale of the heavy right-handed 
Majorana neutrino mass terms providing the well-known see-saw
mechanism of light neutrino mass generation.

In this work, we study in detail the (parton level) discovery potential for the LHC and the Tevatron in the $Z'$ sector of the $B-L$ model, where the $U(1)_{B-L}$ symmetry is gauged, that is not always considered as a traditional benchmark for generic collider reach studies or data analysis.

A very interesting feature of the $B-L$ model is possibly relatively long lifetimes of the heavy neutrinos, which can directly be measured in the decay chain of the $Z'$ boson. Such measurement could also be a key to shedding light on the mass spectra of the light neutrinos.

\section{The model}
The model under study is the so-called ``pure'' or ``minimal''
$B-L$ model (see \cite{bbms} for conventions and references) 
since it has vanishing mixing between the two $U(1)_{Y}$ 
and $U(1)_{B-L}$ groups.
%In the rest of this work we refer to this model simply as the ``$B-L$ model''. 
In this model the classical gauge invariant Lagrangian,
obeying the $SU(3)_C\times SU(2)_L\times U(1)_Y\times U(1)_{B-L}$
gauge symmetry, can be decomposed as usual as $\displaystyle
\mathscr{L}=\mathscr{L}_{YM} + \mathscr{L}_s + \mathscr{L}_f + \mathscr{L}_Y$.
The non-Abelian field strengths in $\mathscr{L}_{YM}$ are the same as in the SM
whereas the Abelian ones can easily be diagonalised\footnote{In general, Abelian field strengths tend to mix and the diagonalisation of the kinetic terms could be complicated. However, in our case just one off-diagonal term appears and a linear $2\times 2$ transformation is sufficient to fulfill our aim.}.
%In such field basis, the covariant derivative is:
%$\displaystyle D_{\mu}\equiv \partial _{\mu} + ig_S T^{\alpha}G_{\mu}^{\phantom{o}\alpha} 
%+ igT^aW_{\mu}^{\phantom{o}a} +ig_1YB_{\mu} +i(\widetilde{g}Y + g_1'Y_{B-L})B'_{\mu}\, .$
The ``pure'' or ``minimal'' $B-L$ model is defined as the model in which
%by the condition $\widetilde{g} = 0$, that implies
 there is no mixing between the $Z'$ and the SM $Z$ gauge bosons.

The fermionic Lagrangian is the same as in the SM, apart from the term associated to RH-neutrinos ($i\overline {\nu _{kR}} \gamma _{\mu}D^{\mu} \nu_{kR}$, where $k$ is the
generation index). The fields' charges are the usual SM and $B-L$ ones (in particular, $B-L = 1/3$ for quarks and $-1$ for leptons).
  The latter charge assignments as well as the introduction of three new
  fermionic  right-handed heavy neutrinos ($\nu_R$) and one
  scalar Higgs ($\chi$, charged $+2$ under $B-L$)  
  fields are designed to eliminate the triangular $B-L$  gauge anomalies and to ensure the gauge invariance of the theory, respectively.
  Therefore, the $B-L$  gauge extension of the SM group
  broken at the Electro-Weak (EW) scale does necessarily require
  at least one new scalar field and three new fermionic fields charged with respect to the $B-L$ group.

\begin{comment}
Regarding the scalar Lagrangian, the only differences with respect to the SM is that we have to introduce a kinetic term for the $\chi$ field as well as to modify the scalar potential, given by
%\begin{equation}\label{new-potential}
$\displaystyle V(H,\chi ) = m^2H^{\dagger}H +
 \mu ^2\mid\chi\mid ^2 +
  \lambda _1 (H^{\dagger}H)^2 +\lambda _2 \mid\chi\mid ^4 + \lambda _3 H^{\dagger}H\mid\chi\mid ^2  \, ,$
%\end{equation}
{where $H$ and $\chi$ are the complex scalar Higgs 
doublet and singlet fields, respectively.}
\end{comment}

Finally, as for Yukawa interactions, we are allowed to introduce two new terms ($\displaystyle y^{\nu}_{jk}\overline {l_{jL}} \nu _{kR}$ $\widetilde H 
	         + y^M_{jk}\overline {(\nu _R)^c_j} \nu _{kR}\chi$, where $\widetilde H=i\sigma^2 H^*$ and  $i,j,k$ take the values $1$ to $3$),
where the first term is the usual Dirac contribution and the second term is the Majorana one.
Neutrino mass eigenstates, obtained after applying the see-saw mechanism, will be called $\nu_l$ and $\nu_h$, being the former the SM-like ones. With a reasonable choice of the Yukawa couplings, the heavy neutrinos $\nu_h$ can have masses $m_{\nu_h} \sim \mathcal{O}(100)$ GeV, within the LHC reach. Their role will be discussed later on.

The model is significantly different from other common $Z'$ models concerning the gauge coupling, $g'_1$, only vectorial and not fixed a priori from unification, and due to the presence of new coupled matter, the heavy neutrinos, providing not fixed $Z'$ boson's branching ratios (BRs).

\section{$Z'$ phenomenology and discovery potential at the LHC and at the Tevatron}
 
In the $B-L$ model, the $Z'$ boson predominantly couples to
leptons. After summing over the generations, leptons roughly get a total BR of $3/4$ and quarks get the remaining $1/4$.
Not surprisingly then, for a relatively light (with respect to the $Z'$ gauge boson) heavy neutrino, the $Z'$ boson's BR into pairs of such particles is relatively high: $\sim 20\% $ (at most, after summing over the generations).
Regarding finally the total $Z'$ boson width, it strongly depends on the new gauge coupling $g'_1$ up to few hundreds GeV for a TeV scale gauge boson.

The process we are interested in in this work is di-lepton production. We define our signal as $pp\rightarrow (\gamma,\,Z,)\,Z'_{B-L}\rightarrow  \ell^+ \ell^-$ ($\ell=e,\,\mu$), i.e., all possible sources together with their mutual interferences, where between parentesis we indicate also the background, i.e., SM Drell-Yan production (including interference).
For both the signal and the background, we have assumed standard
acceptance cuts (for both electrons and muons) at the LHC(Tevatron), 
$
p_T^\ell > 10(18)~{\rm GeV}$ and $ |\eta^\ell|<2.5(1)$ $(\ell=e,\,\mu).
$ 
On the di-lepton invariant mass,  $M_{\ell\ell}$, 
 we will select a window as large as either one width of the $Z'_{B-L}$ boson or twice the di-lepton mass resolution, 
whichever the largest. The half windows in the invariant mass distributions  read for the LHC\footnote{We take  the CMS mass resolutions \cite{CMS-CDFdet} as typical for the LHC environment.} as
$|M_{ee}-M_{Z'}| < 
\mbox{max} \left( \frac{\Gamma_{Z'}}{2},\; \left( 0.02\frac{M_{Z'}}{\rm GeV} \right) {\rm GeV}\; \right)$ for electrons and  $|M_{\mu\mu}-M_{Z'}| < 
\mbox{max} \left( \frac{\Gamma_{Z'}}{2},\; \left( 0.08\frac{M_{Z'}}{\rm GeV} \right) {\rm GeV}\; \right)$ for muons and for the Tevatron\footnote{We take the CDF mass resolutions \cite{CMS-CDFdet} as typical for the Tevatron environment.} as
$|M_{ee}-M_{Z'}| < 
\mbox{max} \left( \frac{\Gamma_{Z'}}{2},\; \left( 0.135 \sqrt{\frac{M_{Z'}}{\rm GeV}}{\rm GeV}+ 0.02\frac{M_{Z'}}{\rm GeV} \right) {\rm GeV}\; \right)$ again for electrons and $|M_{\mu\mu}-M_{Z'}| <
\mbox{max} \left( \frac{\Gamma_{Z'}}{2},\; \left( 0.0005\left(\frac{M_{Z'}}{2\, \rm GeV}\right) ^2 \right) {\rm GeV}\; \right)$ for muons.
Our approach is to count all the events within this window, effectively accounting for the convolution between the two components.
We finally define the \emph{significance} $\sigma$ as $s/ \sqrt{b}$ ($s$ and $b$ being the signal and background event rates, respectively)\footnote{This definition, based on a gaussian distribution, is valid when the number of events is large enough, i.e. $s,\,b>20$. Otherwise, in case of lower statistics, we exploited the Bityukov algorithm \cite{Bityukov}, which uses the Poisson `true' distribution.}: the discovery will be for $\sigma\ge 5$, as usual.

As a result, it is clear that the Tevatron will still be 
competitive with the LHC, especially in the lower mass region where the LHC requires $1$ fb$^{-1}$ to be sensitive to the same couplings as the Tevatron, down to $\sim (3 \div 4) \cdot 10^{-2}$ for electrons and to $\sim (4 \div 5) \cdot 10^{-2}$ for muons. The LHC is not limited by the kinematics, being able to discover the $Z'_{B-L}$ boson up to masses of $1.2$ TeV for $1$~fb$^{-1}$, while at the Tevatron a $5\sigma$ discovery will be possible up to a value of the mass of $0.9$~TeV, as also clear if a study of the required luminosity at fixed coupling is performed. This study shows as well that, for $g'_1=0.1$, the LHC requires $0.25(0.45)$ fb$^{-1}$ to be sensitive at $5\sigma$, while Tevatron requires $5(10)$ fb$^{-1}$, with electrons(muons), respectively.

If no evidence for a signal is found at any configuration, $95\%$~C.L. exclusions limits can be derived, and the limits derived using electrons are always more stringent than those derived using muons due to the different resolutions.
%The different resolutions imply that the limits derived using electrons are always more stringent than those derived using muons.
In particular, at the Tevatron with $10$ fb$^{-1}$, the $Z'_{B-L}$ boson can be excluded for values of the coupling down to $0.03$ ($0.04$ for muons) for $M_{Z'}=600$ GeV. For the LHC to set the same exclusion limit, $1$ fb$^{-1}$ is required. For the same integrated luminosity, the LHC has much more scope in excluding a $Z'_{B-L}$, when $M_{Z'}>800$ GeV, and the exclusion is possible up to $M_{Z'}=1.6$ TeV for $1$ fb$^{-1}$ (both with electrons and muons).

A comparison between the $Z'_{B-L}$ discovery power at the LHC and at a future Linear Collider is found at Ref.~\cite{bbmp}.

\section{Neutrino phenomenology and mass measure}

The possibility of the $Z'$ gauge boson decaying into pairs of heavy
neutrinos is a very peculiar feature of this model,
since, in addition to the clean di-lepton signature, it
provides multi-lepton signatures where backgrounds can strongly be 
suppressed \cite{bbms}. Hence, it deserves to be looked at in details once a $Z'$ boson will eventually be observed, also to disentangle almost uniquely this model.
Regarding the neutrino sector, after the see-saw diagonalisation of the neutrino mass matrix, we obtain three very light neutrinos ($\nu_l$), which are the SM-like neutrinos, and three heavy neutrinos ($\nu_h$).
The latter have an extremely small mixing
with the  $\nu_l$'s thereby providing very small but non-vanishing 
couplings to gauge and Higgs bosons. 
Hence, neglecting the scalar sector, the $\nu_h$'s prefer to decay into SM gauge bosons, as well as into the new $Z'$, when these decay channels are kinematically allowed.
% which in turn enable the $\nu_h$ to dominantly decay into SM gauge bosons (neglecting the scalar sector), as well as into the new $Z'$ when these decay channels are kinematically allowed.
In detail, BR$\left( \nu_h \rightarrow l^\mp W^\pm \right)$
is dominant and reaches the $2/3$ level in the  $M_{Z'}> M_{\nu_h} \gg M_W, M_Z$
limit, while the BR$\left( \nu_h\rightarrow \nu_l Z \right)$ represents the remaining $1/3$ in this regime. 
%In contrast, the $\nu_h \rightarrow \nu_l h_2$ as well as $\nu_h \rightarrow \nu_l Z'$ decay channels are well below the percent level and are negligible.

The $\nu_h$ couplings to the gauge bosons are
proportional to the ratio of light to heavy neutrino masses, which is extremely small.  The decay width of
the heavy neutrino is correspondingly small and its lifetime large: it can be a long lived particle and, over a
large portion of parameter space, its lifetime can be comparable to or
exceed that of the $b$-quark, giving rise to a displaced vertex inside the detector.
The key point is that a measurable lifetime along with a mass
determination for $\nu _h$ also enables a determination of
$m_{\nu _l}$, by just inverting the see-saw formula.

Multi-lepton signatures carry the hallmark of the heavy neutrinos
as the latter enter directly the corresponding decay chains.
%and the further leptons come from the $W$ and $Z$ following decays.
We performed a detailed Monte Carlo analysis
at the benchmark point $M_Z' = 1.5$ TeV, $g'_1 = 0.2$
and $M_{\nu_h}=200$ GeV, with $\sigma (pp\rightarrow \nu _h \nu_h) = 46.7$ fb (for CTEQ6L PDFs with $Q^2=M_{Z'}^2$), and the decay we are interested in is $Z'\rightarrow \nu _h \nu_h \rightarrow 3l+2q+\nu _l$, with a significant fraction of missing energy. A suitable distribution to look at turned out to be the transverse mass defined in \cite{Barger}, i.e.,
%\begin{equation}
$\displaystyle m^2_T = \left( \sqrt{M^2(vis)+P^2_T(vis)}+\left| \mpt \right| \right) ^2
	- \left( \vec{P}_T(vis) + \mptv\right) ^2\, $,
%\end{equation}
where $(vis)$ means {the sum over} the visible particles.  If the
visible particles we sum over are the $3$ leptons and $2$
jets, the transverse mass distribution will peak at the $Z'$ boson mass. We can
also see evidence for the presence of a heavy neutrino by just
considering as visible particles the $2$ leptons with the smallest 
azimuth-rapidity separation, since this is the topology of a $\nu_h$ decay.  The
results show that this transverse mass peak for the heavy neutrino is
likely to be the best way to measure its mass. Also, the backgrounds are well under control after simple kinematic and detector acceptance requirements. The striking signature of
this model is that both of the above peaks occur simultaneously.

We have therefore proven that substantial scope exists at both the Tevatron and LHC in detecting a $Z'_{B-L}$ state in a variety of decay channels.

%It is important to note that the backgrounds are completely under control: as sources we considered $WZjj$, $t\overline{t}$ (where a further lepton comes from a semileptonic $b$ decay) and $t\overline{t}l\nu$. After simple kinematic and detector acceptance requirements, significant suppression of the backgrounds comes from requiring the di-jet invariant mass to be close to the $W$ mass (as this is the signal topology). At this stage, just the $WZjj$ is comparable to our signal, but the $Z'$ peak described above can be used to further reduce the remnant background.

%\section{Conclusions}

\end{document}